# Demonstration of Duality of Fractal Gravity Models by Scaling Symmetry

Yanguang Chen

(Department of Geography, College of Urban and Environmental Sciences, Peking University, Beijing 100871, P.R. China. E-mail: chenyg@pku.edu.cn)

**Abstract**: A pair of fractal gravity models can be derived from the spatial interaction models based on entropy maximizing principle and allometric scaling law. The models can be expressed as dual form in mathematics and are important for analyzing and predicting spatial flows in network of cities. However, the dual relationship of urban gravity parameters has been an empirical relationship for a long time and lacks theoretical proof. This paper is devoted to proving the duality of fractal gravity models by means of ideas from scaling invariance. The results show that a fractal gravity model can be derived from its dual form. The observational data of interurban spatial flows of Beijing-Tianjin-Hebei region of China are employed to make a case study, lending a further support to the theoretical derivation. A conclusion can be reached that the duality of gravity models rests with scaling symmetry of fractal structure. This work may be helpful for understanding the theoretical essence and application direction of spatial interaction modeling.

## 1. Introduction

Cities are complex systems and spatial interaction reflects complex spatial dynamics of urban evolution. Interaction is one of the ubiquitous process across the individual sciences (Liu *et al*, 2003). The process cannot be effectively understood within the set of references developed within the specific scientific fields. Urban dynamics provides a spatial perspective for understanding the interaction process. Cities used to be understood as places in space. Today, this understanding is not enough for us to making scientific research on cities. I quite agree with Batty (2013) who argued that cities are not simply as places in space but as systems of networks and flows. To understand space, we must understand intra-urban and interurban flows, and to understand flows, we must understand networks, and the relations between objects that comprise the city system (Batty, 2013). Fractal gravity models can play an important role in spatial analysis and prediction of flows in a network of cities. The models bears a set of scaling exponent and take on a pair of dual forms, and the cross scaling exponents can be associated with fractal parameters (Chen, 2015). Though, the fractal gravity models can be derived from the spatial interaction models developed by Wilson (1968, 1970, 2000, 2010), the dual relationship between the two expressions of the gravity models are still empirical relationships and lack theoretical proof. Owing to absence of theoretical demonstration, urban scientists cannot always believe the effect of the dual expression of the gravity model in geography.

The theoretical proof of the dual relationship between the two expressions of gravity models is necessary and possible today. This paper is devoted to deriving the dual expression of a gravity model from its original form. The method is mathematical reasoning based on scaling transformation. The aim is to change the empirical dual relation into a theoretical symmetric relations. The rest parts are arranged as follows. In Section 2, the dual relationships of gravity models based on inverse power-law decay function are mathematically proved by means of ideas from scaling symmetry. In Section 3, based on interurban passenger flows and nighttime light data of cities in Beijing-Tianjin-Hebei region, China, am empirical analysis is made to lend a further support to the theoretical results. In Sections 4 and 5, the mathematically proved results for power-law based gravity models is generalized to the dual relationships of the gravity models based on negative exponential decay function, several related questions are discussed, and finally, the discussion is concluded with summarizing the main points of this study.

## 2. Models

### 2.1 Duality of gravity models

One of basic measurement of gravity in urban geography is spatial flows, including passenger flow, cargo flow, and information flow. By means of allometric scaling law, a generalized gravity models of spatial flow can be derived from Wilson's spatial interaction model (Chen, 2015). Based on outflow, the model can be expressed as follows

$$T_{ij} = KP_i^\mu P_j^\nu r_{ij}^{-\sigma} ,\qquad(1)$$

where $T_{ij}$ denotes the outflow quantity from city $i$ to city $j$, $P_i$ is the size of city $i$, and $P_j$ the size of city $j$, $r_{ij}$ refers to the distance between city $i$ and city $j$. As for the parameters, $K$ is proportionality coefficient, $u$ and $v$ refer to size exponents, and $\sigma$ is distance decay exponent. The size exponents are also termed "adjustable exponent" (Simini *et al*, 2012). All these exponents are actually cross scaling exponents. If $u=v=1$ as given, we will have standard form of gravity model (Batty and Karmeshu, 1983). However, in empirical analysis, $u$ is not equal to $v$. Therefore, the standard gravity model needs new definition. Based on inflow, the model can be expressed as below:

$$T_{ji} = KP_i^\nu P_j^\mu r_{ji}^{-\sigma} ,\qquad(2)$$

where $T_{ji}$ denotes the inflow quantity from city $j$ to city $i$. Apparently, there is a dual relationship between equation (1) and equation (2). If equation (1) is regarded as the original form, then equation (2) represents the dual form and *vice versa*.

The dual gravity models are useful in geo-spatial analysis of network and flows. In fact, all mathematical modelling have two major aims: explanation and prediction (Batty, 1991; Fotheringham and O'Kelly, 1989; Kac, 1969). The dual gravity models can be employed to characterize spatial flows, estimate fractal dimension of traffic networks, predict spatial flow quantity and make up missing data, reveal the problems in a traffic network by analyzing prediction residuals, especially, outliers of residuals, and so on. In theory, the dual relationship is an odd symmetric relationship. The relationship has been testified by observational data (Chen, 2015). However, so far, it is only an empirical relationship, lacking theoretical proof.

## 2.2 Duality proof by means of scaling symmetry

In terms of scaling symmetry principle, a mathematic proof can be made for the dual relationships of the gravity models. For a given urban system, inflow quantity is generally different from outflow quantity. However, due to the symmetry of distance matrix, the distance decay exponent value based on inflow quantity is equal to that based on outflow quantity. For convenience of the proof, equation (2) is rewritten as follows

$$T_{ji} = KP_i^\alpha P_j^\beta r_{ij}^{-\sigma}, \tag{3}$$

where $\alpha$ and $\beta$ to size scaling exponents. If we can prove that $u=\beta$ and $v=\alpha$, the problem will be solved. Due to asymmetry of inflow and outflow, we have $T_{ij} \neq T_{ji}$. However, the total flow quantity is certain, that is

$$T = \sum_{i=1}^n \sum_{j=1}^n T_{ij} = \sum_{i=j}^n \sum_{i=1}^n T_{ji} . \tag{4}$$

A square matrix multiplied by its transposed form is a symmetric matrix. Equation (1) multiplied by equation (3) yields

$$T_{ij}T_{ji} = K^2 P_i^{u+\alpha} P_j^{v+\beta} r_{ij}^{-2\sigma} . \tag{5}$$

On the other hand, the matrix formed by the product of inflow and outflow must be symmetric, that is

$$T_{ji}T_{ij} = K^2 P_j^{u+\alpha} P_i^{v+\beta} r_{ji}^{-2\sigma} . \tag{6}$$

Thus we have

$$P_i^{u+\alpha} P_j^{v+\beta} = P_j^{u+\alpha} P_i^{v+\beta} . \tag{7}$$

So, there must be

$$u + \alpha = v + \beta . \tag{8}$$

On the other hand, the double summation of equations (1) and (3) are as follows

$$T = \sum_{i=1}^n \sum_{j=1}^n T_{ij} = K \sum_{i=1}^n \sum_{j=1}^n P_i^u P_j^v r_{ij}^{-\sigma} , \tag{9}$$

$$T = \sum_{i=1}^n \sum_{j=1}^n T_{ji} = K \sum_{i=1}^n \sum_{j=1}^n P_j^\beta P_i^\alpha r_{ji}^{-\sigma} . \tag{10}$$

Making scaling transformation of equations (9) and (10) yields

$$T(\gamma P_i, \gamma P_j) = \gamma^{u+v} K \sum_{i=1}^n \sum_{j=1}^n P_i^u P_j^v r_{ij}^{-\sigma} = \gamma^{u+v} T(P_i, P_j) = \lambda T(P_i, P_j) , \tag{11}$$

$$T(\gamma P_i, \gamma P_j) = \gamma^{\alpha+\beta} K \sum_{i=1}^n \sum_{j=1}^n P_i^\alpha P_j^\beta r_{ij}^{-\sigma} = \gamma^{\alpha+\beta} T(P_i, P_j) = \lambda T(P_i, P_j), \tag{12}$$

where $\gamma$ denotes the scale factor for scaling transformation, and $\lambda$ refers to eigenvalue of the scaling transform. Equations (11) and (12) are based on the idea from scaling symmetry of power law decay. Since the total flow is constant, there must be

$$\gamma^{u+v} = \gamma^{\alpha+\beta} = \lambda. \tag{13}$$

This gives the following parameter relation

$$u + v = \alpha + \beta. \tag{14}$$

Equation (8) plus equation (14) yields

$$u = \beta. \tag{15}$$

Equation (8) minus equation (14) yields

$$v = \alpha. \tag{16}$$

This implies that equation (3) is equivalent to equation (2). The proof is complete.

## 2.3 Pure gravity model and normal distance exponent

Spatial flow quantity can be employed to measure gravity, but spatial flow itself is not attractive force. The dual gravity models are actually a pair of dual flow models and can be transformed into standard form of gravity model. If so, the model parameters, including gravity coefficient $K$ and distance exponent $\sigma$ will be standardized. The production of equations (1) and (2) is

$$T_{ij}T_{ji} = K^2 P_i^{u+v} P_j^{v+u} r_{ij}^{-2\sigma}. \tag{17}$$

which can be converted into the standard form of gravity model as below

$$I_{ij} = (T_{ij}T_{ji})^{1/(u+v)} = K^{2/(u+v)} P_i P_j r_{ij}^{-2\sigma/(u+v)} = GP_i P_j r_{ij}^{-b} = I_{ji}, \tag{18}$$

in which the model parameters are as follows

$$G = K^{2/(u+v)}, \tag{19}$$

$$b = \frac{2\sigma}{u+v}. \tag{20}$$

The two formulae represent the normal forms of gravity coefficient and distance exponent. It can be proved that equation (18) is a fractal models, and thus equations (1) and (2) represents a pair of scaling models for flows and gravity.

The distance decay exponent of the standard gravity model proved to be associated with Zipf's and central place fractals. Both Zipf's law and central place fractals bear scaling symmetry (Batty and Longley, 1994). This suggests the symmetry behind the gravity models. Based on the self-similar hierarchy of central place networks, a parameter equation can be derived as follows

$$b = qD, \tag{21}$$

where $b$ is the distance exponent, $q$ is the rank-size scaling exponent of Zipf's distribution, and $D$ denotes the fractal dimension of central place networks (Chen, 2015). The textural fractals of central place models were revealed by Arlinghaus (1985, 1993) and her co-workers (Arlinghaus and Arlinghaus, 1989). From textural central place fractals, we can derive structural central place fractals of central place models. In theory, the Zipf exponent $q$ approaches to 1, that is, $q \rightarrow 1$, and central place fractal dimension comes between 1 and 2. Therefore, we have

$$u + v \leq \sigma .$$ 

(22)

For the standard central place network, the dimension $D=d=2$. Therefore, in the extreme condition, the distance exponent $b \rightarrow 2$. Under these circumstances, we have $u+v=\sigma$. Of course, this is a special case. But for the reasonable structure of central place system, the distance exponent varies from 1 to 2. However, the fractal dimension $D$ corresponds to similarity dimension rather than box dimension. This suggests that the fractal dimension, and thus the distance exponent, may depart significantly from the proper interval owing to random disturbance in the real world.

## 2.4 Analogy and extension

The proved results of dual gravity models based on inverse power law decay can be generalized to the models based on negative exponential decay. Based on translational symmetry of distance, equations (1) and (2) can be revised as

$$T_{ij} = KP_i^u P_j^v e^{-r_{ij}/r_0} ,$$

(23)

$$T_{ji} = KP_i^v P_j^u e^{-r_{ji}/r_0} ,$$

(24)

where $r_0$ denotes the characteristic distance of spatial flows, the other symbols are the same as those in equations (1) and (2). In this models, city sizes follows hierarchical scaling law, while interurban distances does not follow spatial scaling law. Hierarchical scaling law is based on allometric growth law and Zipf's law (Chen, 2011; Chen, 2015). Imitating the proof process in Subsection 2.2, we can deduce equation (24) from equation (23) and *vice versa*. This is because that the above scaling transformation only involves the sizes of cities, not the distances between cities. The product of equations (23) and (24) is as below:

$$T_{ij} T_{ji} = K^2 P_i^{u+v} P_j^{v+u} e^{-2r_{ij}/r_0} ,$$

(25)

which can be transformed into standard form as below:

$$I_{ij} = (T_{ij}T_{ji})^{1/(u+v)} = K^{2/(u+v)}P_iP_je^{-2r_{ij}/(r_0(u+v))} = GP_iP_je^{-r_{ij}/r_0^*} = I_{ji} \,, \qquad (26)$$

in which the rescaled characteristic distance is as follows

$$r_0^* = \frac{(u+v)r_0}{2} = \frac{(u+v)\overline{d}}{4} \,, \qquad (27)$$

where $d$ bar denotes the average distance. This suggests that the characteristic distance $r_0$ is the half of the average distance. The notation of gravity coefficient is the same as that of the gravity model based on power law decay, that is, equation (19). A power law decay suggests long-range interaction, while an exponent law suggest spatial locality. The former indicates scaling symmetry, while the latter implies translational symmetry. In this sense, the power-law distance decay function is more consistent with the first law of geography.

# 3. Empirical evidence

## 3.1 Data and methods

The effect of a mathematical model needs to be evaluated by observation data. If a mathematical model reflects a regularity, it can be supported by the observed data. If a mathematical derivation is valid, it can also be verified by the observed data. I quite agree with Louf and Barthelemy (2014), who said: "The success of natural sciences lies in their great emphasis on the role of quantifiable data and their interplay with models. Data and models are both necessary for the progress of our understanding: data generate stylized facts and put constraints on models. Models on the other hand are essential to comprehend the processes at play and how the system works. If either is missing, our understanding and explanation of a phenomenon are questionable. This issue is very general, and affects all scientific domains, including the study of cities." The study area is Beijing-Tianjin-Hebei region of China. The observational data bear three sources. The spatial flows of passengers were mined from Baidu migration platform, the distances were extracted by ArcGIS, and the city sizes were measured by urban nighttime light area and total quantity (Table 1). The data of area and total numbers of urban nighttime light are defined within built-up area of cities in Beijing-Tianjin-Hebei region.

**Table 1 The measures and data source for empirical analysis of dual gravity models**

| Measure | Symbol | Meaning | Data source | Year |
|---------|--------|---------|-------------|------|
| **Flow** | $T_{ij}$ | Regional population flow | Baidu migration platform | 2017 |
| **Distance** | $r_{ij}$ | Interurban distance | Extraction by ArcGIS | 2017 |
| **City size** | $P_i, P_j$ | Nighttime light area and number | American NOAA National Centers for Environmental Information (NCEI) | 2016 |

**Note**: Baidu migration platform (http://qianxi.baidu.com/).

The multivariate linear regression based on least squares method (LSM) can be employed to estimate models' parameter values. By taking logarithms, the dual gravity models can be turned into linear forms. Equation (17) can be turned into the following form

$$\ln(\sqrt{T_{ij}T_{ji}}) = \ln K + \frac{u+v}{2}\ln P_i + \frac{v+u}{2}\ln P_j - \sigma \ln r_{ij}. \tag{28}$$

In fact, equations (1) and (2) can be transformed into the following expressions

$$\ln T_{ij} = \ln K + u \ln P_i + v \ln P_j - \sigma \ln r_{ij}, \tag{29}$$

$$\ln T_{ji} = \ln K + v \ln P_i + u \ln P_j - \sigma \ln r_{ji}. \tag{30}$$

Then, by means of least squares calculation, the model parameters can be worked out easily (Tables 2 and 3). The parameter values make significant support to the theoretical derivation.

## 3.2 Results

Different forms of the same model are suitable for different research directions. If we want to generate a standard gravity model, we can use equation (28) to estimate model parameters. If we try to prove the duality of the inflow gravity model and outflow gravity model, equations (29) and (30) can give the direct and visual results of parameter estimation. First of all, the nighttime light areas within built-up area of cities are used as city size measurements for gravity modeling. The results of least square regression for parameters estimation are displayed in Table 2. Based on the outflow, the gravity model is

$$\hat{T}_{ij} = 2180064.7343 P_i^{0.7851} P_j^{0.4381} r_{ij}^{-2.0026}, \tag{31}$$

The goodness of fit is $R^2$=0.6686. The parameter values were estimated by using equation (29). Based on the inflow, the gravity model is

$$\hat{T}_{ji} = 2180064.7343 P_i^{0.4381} P_j^{0.7851} r_{ji}^{-2.0026}, \tag{32}$$

The goodness of fit is $R^2$=0.6686. The parameter values were estimated by using equation (30). Apparently, there is a dual relationship between the cross scaling exponents of equation (31) and those of equation (32). Based on the geometric mean of inflow and outflow, the gravity model is

$$\sqrt{\hat{T}_{ij}\hat{T}_{ji}} = 2180064.7343 P_i^{0.6116} P_j^{0.6116} r_{ji}^{-2.0026}. \tag{33}$$

The goodness of fit is $R^2$=0.7169. The parameter values were estimated by using equation (28). Based on equation (31), (32), or (33), a standard gravity model can be given as follows

$$\hat{I}_{ij} = 23146308755.7504 P_i P_j r_{ij}^{-3.2746}. \tag{34}$$

The distance decay exponent is $b$=3.2746, which departs from the reasonable interval of fractal dimension.

**Table 2 The estimated parameter values of dual gravity models and related statistics based on urban nighttime light area**

| Type | Name | Based on outflows | | Based on inflows | |
|---|---|---|---|---|---|
| **Global statistics** | Multiple $R$ | 0.8177 | | 0.8177 | |
| | $R$ Square | 0.6686 | | 0.6686 | |
| | Adjusted $R$ Square | 0.6614 | | 0.6614 | |
| | Standard error | 1.0097 | | 1.0097 | |
| | Number of observations | 142 | | 142 | |
| | Analysis of variance | $F$ | Significance $F$ | $F$ | Significance $F$ |
| | | 92.8156 | 6.1716E-33 | 92.8156 | 6.1716E-33 |
| **Parameter and local statistics** | | Coefficients | $P$-value | Coefficients | $P$-value |
| | Intercept | ln$K$=14.5949 | 1.2569E-22 | ln$K$=14.5949 | 1.2569E-22 |
| | ln$P_i$ | $u$=0.7851 | 6.9397E-17 | $v$=0.4381 | 3.9986E-07 |
| | ln$P_j$ | $v$=0.4381 | 3.9986E-07 | $u$=0.7851 | 6.9397E-17 |
| | ln$r_{ij}$ | $\sigma$=-2.0026 | 4.5684E-23 | $\sigma$=-2.0026 | 4.5684E-23 |

**Note**: The coefficients represent the regression parameters of equations (29) and (30). The dual relationship between the inflow gravity model and outflow gravity model can be reflected by the coefficient values. The probability value of parameters are equivalent to corresponding student's t statistics. These notes are suitable for Table 3.

Then, the total number of nighttime lights within built-up area of cities served as city size measurements for gravity modeling. The results of least square calculations for parameters estimation are listed in Table 3. Based on the outflow, the gravity model is

$$\hat{T}_{ij} = 135852.8300 P_i^{0.6624} P_j^{0.3117} r_{ij}^{-1.9824}, \tag{35}$$

The goodness of fit is $R^2$=0.6547. The parameter values were estimated by means of equation (29). Based on the inflow, the gravity model is

$$\hat{T}_{ji} = 135852.8300 P_i^{0.3117} P_j^{6624} r_{ji}^{-1.9824}, \tag{36}$$

The goodness of fit is $R^2$=0.6547. The parameter values were estimated by means of equation (30). Obviously, there is a dual relationship between the cross scaling exponents of equation (35) and those of equation (36). Based on the geometric mean of inflow and outflow, the gravity model is

$$\sqrt{\hat{T}_{ij}\hat{T}_{ji}} = 135852.8300 P_i^{0.4870} P_j^{4870} r_{ji}^{-1.9824}. \tag{37}$$

The goodness of fit is $R^2$=0.6887. The parameter values were estimated by means of equation (28). From equation (35), (36), or (37), a standard form of gravity model can be derived as

$$\hat{I}_{ij} = 34609142509.3512 P_i P_j r_{ij}^{-4.0702}. \tag{38}$$

The distance decay exponent is $b$=4.0702, which also departs from the reasonable interval of fractal dimension.

**Table 3 The estimated parameter values of dual gravity models and related statistics based on total numbers of urban nighttime lights**

| Type | Name | Based on outflows | | Based on inflows | |
|---|---|---|---|---|---|
| **Global statistics** | Multiple $R$ | 0.8091 | | 0.8091 | |
| | $R$ Square | 0.6547 | | 0.6547 | |
| | Adjusted $R$ Square | 0.6471 | | 0.6471 | |
| | Standard error | 1.0308 | | 1.0308 | |
| | Number of observations | 142 | | 142 | |
| | Analysis of variance | $F$ | Significance $F$ | $F$ | Significance $F$ |
| | | 87.2010 | 1.0547E-31 | 87.2010 | 1.0547E-31 |
| **Parameter and local statistics** | | Coefficients | $P$-value | Coefficients | $P$-value |
| | Intercept | ln$K$=11.8193 | 1.4477E-12 | ln$K$=11.8193 | 1.4477E-12 |
| | ln$P_i$ | $u$=0.6624 | 1.4539E-16 | $v$=0.3117 | 1.9060E-05 |
| | ln$P_j$ | $v$=0.3117 | 1.9060E-05 | $u$=0.6624 | 1.4539E-16 |
| | ln$r_{ij}$ | $\sigma$=-1.9824 | 4.4621E-22 | $\sigma$=-1.9824 | 4.4621E-22 |

All the calculation results lend support to the theoretical inference that the cross scaling exponents of the inflow gravity model is dual to those of the outflow gravity model. Based on the nightlight area of the cities in Beijing-Tianjin-Hebei region, the parameter is as follows: $u$=0.7851, $v$=0.4381. The other parameter values and related statistics of the outflow gravity model are the same as those of the inflow model (Table 2). Based on the nighttime light number of the cities in the study, the parameter is as follows: $u$=0.6624, $v$=0.3117. The other parameter values and related statistics of the inflow gravity model are also equal to those of the outflow model (Table 3).

For comparison, the gravity model based on negative exponential decay function may be used to fit the observed data. The results are tabulated as follows (Table 4). If the nighttime light area is used as size measurement, the goodness of fit for gravity models based on negative exponential decay function is better than that of the gravity models based on inverse power law decay. In contrast, if the nighttime light number is used as size measurement, the goodness of fit for gravity models based on inverse power law decay is better than that of the gravity models based on negative exponential decay function. Models selection and evaluation is not the topic of this work. It can be seen that the model parameters verify the dual relationships between the inflow gravity model and outflow gravity model based on exponential decay law.

**Table 4 The dual gravity models and corresponding goodness of fit based on negative exponential decay function**

| Size measure | Type | Exponential Gravity Model | Goodness of fit |
|---|---|---|---|
| **Nighttime light area** | Outflow | $\hat{T}_{ij} = 308.3268 P_i^{0.7881} P_j^{0.4411} e^{-r_{ij}/126.1662}$ | 0.6701 |
| | Inflow | $\hat{T}_{ji} = 308.3268 P_i^{0.4411} P_j^{0.7881} e^{-r_{ij}/126.1662}$ | 0.6701 |
| | Double flow | $\sqrt{\hat{T}_{ij}\hat{T}_{ji}} = 308.3268 P_i^{0.6146} P_j^{0.6146} e^{-r_{ij}/126.1662}$ | 0.7185 |
| | Gravity | $\hat{I}_{ij} = 11208.0586 P_i P_j e^{-r_{ij}/77.5470}$ | 0.7185 |
| **Nighttime light number** | Outflow | $\hat{T}_{ij} = 23.3533 P_i^{0.6581} P_j^{0.3073} e^{-r_{ij}/128.5817}$ | 0.6493 |
| | Inflow | $\hat{T}_{ji} = 23.3533 P_i^{0.3073} P_j^{0.6581} e^{-r_{ij}/128.5817}$ | 0.6393 |
| | Double flow | $\sqrt{\hat{T}_{ij}\hat{T}_{ji}} = 23.3533 P_i^{0.4827} P_j^{0.4827} e^{-r_{ij}/128.5817}$ | 0.6827 |
| | Gravity | $\hat{I}_{ij} = 683.4898 P_i P_j e^{-r_{ij}/62.0675}$ | 0.6827 |

The fractal nature of spatial flows is theoretically associated with the fractal structure of road networks. Traffic and transportation networks have been proved to bear fractal nature (Chen *et al*, 2019; Frankhauser, 1990; Kim *et al*, 2003; Lu and Tang, 2004; Sun, 2007). Railway networks show fractal property (Benguigui and Daoud, 1991; Valério *et al*, 2016). Particularly, a great many studies demonstrate that urban road and street network can be describe with fractal geometry and analyzed by using fractal parameters (Bai and Cai, 2008; Chen and Long, 2021; Lu *et al*, 2016; Prada *et al*, 2019; Rodin and Rodina, 2000; Sahitya and Prasad, 2020; Sun *et al*, 2012; Wang *et al*, 2017). If the impedance function of a gravity model departs from inverse power law, the macro patterns of spatial flows deviates the fractal structure. One the other, if the distance decay function follows a power law, but the distance exponent goes beyond the proper interval for fractal dimension, the micro interaction of elements deviates fractal mode. The abnormal values of parameters of gravity models suggest the problem in spatial dynamics. These questions are interesting, but they are not the subject of this work.

## 4. Discussion

Mathematical models including theoretical models and empirical models. The former is also termed mechanism models, while the latter is also termed parameters models (Su, 1988). Where the approaches to making models is concerned, mechanism models are based on analytical method, while parameters models is based on experimental method (Zhao and Zhan, 1991). The analytical process for modeling is based on exploring a system's structure, while the experimental process for modeling is based on testing a system's function. Thus it can be seen that mechanism models can be regarded as structural models and parameter models can regarded as functional models. Wilson's spatial interaction models belong to first category, while the traditional geographical gravity models proceeding from analogy with Newtonian law of universal gravitation belong to the second category (Table 5). The relationship between the two types of models is not invariable. If an empirical model can be mathematically derived from a general principle, or the empirical relation can be proved in mathematical method, the model will change into a theoretical model. Today, the dual gravity model can be treated as a theoretical model rather than an empirical model to a great extent. On the one hand, the general form of the gravity model can be derived from Wilson's spatial interaction model based on entropy maximizing principle (Chen, 2015), on the other hand, the dual relation between the two expressions can be mathematically proved by the ideas from scaling symmetry. A legacy problem is that the logarithmic relationship between transportation cost and distance has not been derived from the principle of entropy maximization.

**Table 5 A comparison between theoretical modeling and empirical modeling in geographical analysis**

| Type | Theoretical models | Empirical models |
|---|---|---|
| **Method of modeling** | Analytical method | Experimental method |
| **Classification of models** | Mechanism models (Structural models) | Parametric models (Functional models) |
| **Scientific world** | Mathematical world | Real world |
| **Sphere of application** | Normative research + value research | Behavioral research+ value research |
| **Case 1** | Wilson's spatial interaction model | Traditional gravity models |
| **Case 2** | Central place models | Zipf's rank-size distribution model |

It is necessary to draw a comparison between the dual gravity models and Wilson's spatial interaction models. Revealing the similarities and differences between the dual gravity models and Wilson's spatial interaction models is helpful for understanding the novelty of the dual gravity models (Table 6). Wilson's spatial interaction model was derived from the principle of entropy maximization (Wilson, 1968; Wilson, 1970). The derivation is actually a nonlinear programming of traffic flows, and the aim is optimize the spatial connections between different regions or cities and make the transportation cost least. In contrast, the gravity models came from analogy with Newtonian law of gravitation (Carey, 1858; Ravenstein, 1885). It was originally proposed as an empirical model rather than a theoretical model (Grigg, 1977). Social scientific research can be divided into three categories: Behavioral research (past + present), normative research (present + future), and values research (past + present + future) (Krone, 1980). Behavioral research is actually positive research or empirical analysis. In contrast, normative research and values research belong to theoretical fields. Therefore, economics were divided into positive economics and normative economics (Behravesh, 2008). Similarly, other social sciences such as human geography should also be divided into positive research and normative research (Chen, 2016).

**Table 6 The similarities and differences between dual gravity models and Wilson's spatial interaction models**

| Type | Dual gravity models | Wilson's spatial interaction models |
|---|---|---|
| **Theoretical basis** | Entropy maximization and allometric scaling law | Entropy maximization and nonlinear programming |
| **Geographical world** | Ideal world | Real world |
| **Distance decay function** | Inverse power law | Negative exponential function |
| **Spatial effect** | Action at a distance | Quasi-locality |
| **Autocorrelation function** | Tail off | Tail off |
| **Partial autocorrelation function** | Tail off | Cut off |
| **Application direction** | Behavioral research and positive analysis | Normative research and predictive analysis |
| **Alternative form** | Based on negative exponential decay | Based on inverse power law decay |

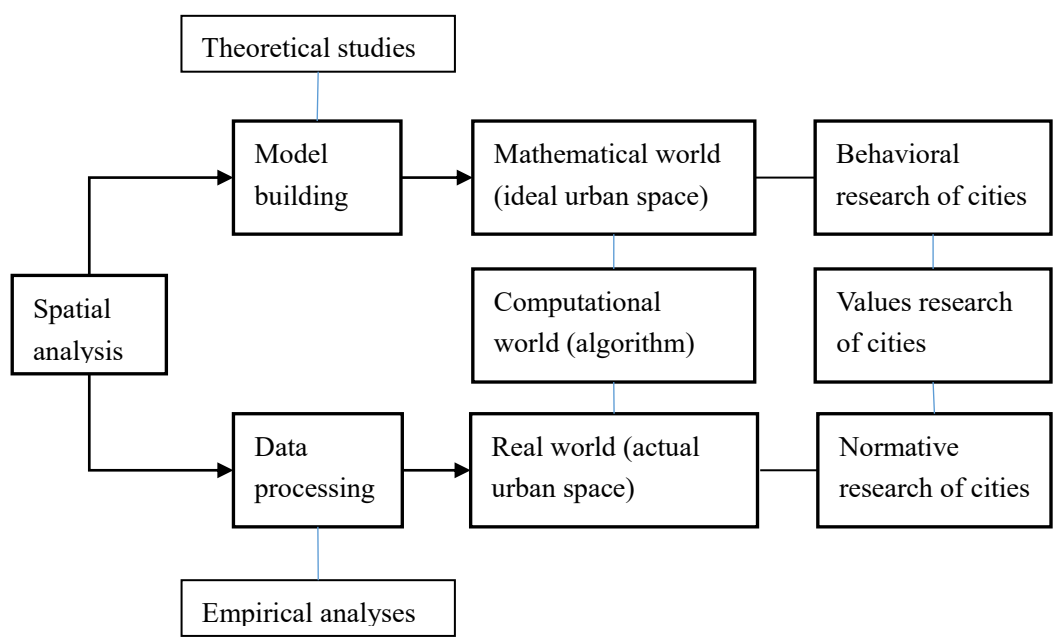

**Figure 1 The geo-spatial analysis based on gravity and spatial interaction modeling can be divided into two directions and three categories**

The three types of geographical research corresponds to the two worlds of geographic analysis and the three worlds of scientific research. In fact, there exists two worlds in geography: real world and ideal world (Tang, 2009). The two geographical worlds corresponds to Plato's empirical world and conceptual world, which are also termed visible world and invisible world (Miller, 1904). The gaps between the two worlds, real world and ideal world, can be treated as the problems to be solved in applied geography. On the other hand, the parameter estimation of any model needs certain algorithms. Application of a mathematical methods to concrete problems involves what is called three worlds: real world, mathematical world, and computational world (Casti, 1996). The computational world can be treated as the bridge between the mathematical world and the real world (Figure 1). The studies on city and regional planning provides the approaches to solve the problems and shorten the distance between the two worlds. In this sense, the dual gravity models is current oriented and based on the past and the real world, while Wilson's spatial interaction models is future oriented and based on the present and the ideal world. The former can be used to make behavioral research and the latter can be utilized to make normative research on networks and flows in geographical world. In empirical research, the dual gravity models can be combined with Wilson's spatial interaction models. It is hard to objectively determine the distance decay function and exponent value for Wilson's spatial interaction. Based on observational data, the form of distance decay function and distance exponent can be judged by using the dual gravity model. Then, based on the results of the distance decay function and exponent from the dual gravity model, Wilson's models can be effectively utilized to predict spatial flows of urban networks. If there are small missed data in a dataset, the dual gravity models can help Wilson's model make up the missing data. In practical work, comparing the difference between the predicted flows of the dual gravity model and those Wilson's spatial interaction models can reveal the problem to be solved in city and regional planning.

The novelty of this work is to provide a theoretical proof for the dual relationships of gravity models by using the ideas of scaling symmetry. What is more, a systemic case study for global gravity analysis is presented to lend more empirical support to the theoretical results. The previous studies based on the general form of gravity model such as equation (1) is mainly for empirical analysis or local interaction analysis (Mackay, 1958; Masucci *et al*, 2013; Batty and Karmeshu, 1983). One of my previous theoretical research showed that the gravity models can be derived from Wilson's spatial interaction models (Chen, 2015). The implicit fractal property behind the models was revealed. The shortcoming of this studies lies in three aspects. First, incomplete spatial dataset.

There are a few missing data in the population mobility data sheet. Second, hybrid data. Railway passenger flows and highway passenger flows were mixed into a dataset. Third, special study period. Special period for data extraction. Population mobility is limited to around the Spring Festival of China. During this special period, the population flow is fast and unstable. Therefore, the standard distance exponent values depart from the proper interval for fractal parameter.

## 5. Conclusions

Spatial interaction modeling is one of important method for spatial analysis. Another significant method is spatial autocorrelation and auto-regression analyses. Gravity models can be treated as the basic tools of spatial interaction analysis. Many forms of gravity models belong to empirical models. After strict mathematical derivation, an empirical model can be sublimated into a theoretical model. This paper makes partial contribution to upgrade the gravity model from empirical model to theoretical model. Based on theoretical proof and empirical results, conclusions can be reached as follows. **First, the duality of the gravity models for outflows and inflows bear strict mathematical foundation.** Duo to the mathematical derivation, the dual relationship is no longer an empirical relationship, but a theoretical relationship. An empirical model is based on observational data, while a theoretical model is based on mathematical derivation or proof and can be supported by observational data. **Second, the dual relationship is not only suitable for the gravity models based on inverse power law decay, but also suitable for those based on negative exponential decay.** Inverse power law decay suggests action at a distance, while negative exponential decay indicates local action of cities. The former is suitable for large scale or complex spatial systems, while the latter is suitable for relatively small scale or simple spatial systems. **Third, the duality of gravity is based on the scaling symmetry of city sizes in the models and symmetry of distance matrix**. One of basic properties of distance matrix is symmetry, which is well known for scientists. The key is scaling symmetry of city sizes in general gravity models based on spatial flows. It is easy to understand the scaling symmetry since the gravity models are associated with rank-size distribution and central place network.

### Acknowledgements

This research was sponsored by the National Natural Science Foundation of China (Grant No.

42171192). The supports are gratefully acknowledged.